\documentclass[runningheads]{llncs}
\usepackage{mathtools}
\usepackage{graphicx}
\usepackage{cite}
\usepackage{makecell}
\usepackage{multirow}
\usepackage{hyperref}
\usepackage{xcolor}
\usepackage{amsmath}
\usepackage{amssymb}
\begin{document}
%
% \title{Contribution Title\thanks{Supported by organization x.}}
\title{Metastatic Cancer Outcome Prediction with Injective Multiple Instance Pooling}
\titlerunning{Metastatic Cancer Outcome Prediction with Injective MIL Pooling}
\author{Jianan Chen 
\inst{1,2}\and  %\orcidID{0000-1111-2222-3333} 
Anne L. Martel\inst{1,2}
}
% \author{Anonymous}
% index{Chen, Jianan}
% index{Milot, Laurent}
% index{Cheung, Helen}
% index{Martel, Anne}
\authorrunning{Anonymous}
% First names are abbreviated in the running head.
% If there are more than two authors, 'et al.' is used.

\institute{
Department of Medical Biophysics, University of Toronto, Toronto, ON, CA  
\\
\email{geoff.chen@mail.utoronto.ca} \and
Sunnybrook Research Institute, Toronto, ON, CA}
% \institute{Anonymous}
% If the paper title is too long for the running head, you can set
% an abbreviated paper title here
%
% \author{Anonymous}
% \author{Jianan Chen\inst{1}\orcidID{0000-1111-2222-3333} \and
% Second Author\inst{2,3}\orcidID{1111-2222-3333-4444} \and
% Third Author\inst{3}\orcidID{2222--3333-4444-5555}}
%
% \authorrunning{F. Author et al.}
% \authorrunning{Anonymous}
% First names are abbreviated in the running head.
% If there are more than two authors, 'et al.' is used.
%
% \institute{Anonymous}
%
\maketitle              % typeset the header of the contribution

\begin{abstract}
Cancer stage is a large determinant of patient prognosis and management in many cancer types, and is often assessed using medical imaging modalities, such as CT and MRI. These medical images contain rich information that can be explored to stratify patients within each stage group to further improve prognostic algorithms. Although the majority of cancer deaths result from metastatic and multifocal disease, building imaging biomarkers for patients with multiple tumors has been a challenging task due to the lack of annotated datasets and standard study framework. In this paper, we process two public datasets to set up a benchmark cohort of 341 patient in total for studying outcome prediction of multifocal metastatic cancer. We identify the lack of expressiveness in common multiple instance classification networks and propose two injective multiple instance pooling functions that are better suited to outcome prediction. Our results show that multiple instance learning with injective pooling functions can achieve state-of-the-art performance in the non-small-cell lung cancer CT and head and neck CT outcome prediction benchmarking tasks. We will release the processed multifocal datasets, our code and the intermediate files \textit{i.e.} extracted radiomic features to support further transparent and reproducible research.
\end{abstract}
\section{Introduction}

The majority of cancer deaths result from metastatic disease \cite{dillekaas201990}. Cancer staging based on the size, number and spread of tumors is one of the predominant tools in the clinic for patient prognosis and management for many tumor types \cite{amin2017eighth}. Medical imaging, such as Computed Tomography (CT) and Magnetic Resonance Imaging (MRI), are often involved in patient standard of care for staging and treatment planning. Beyond their current clinical use, medical images contain additional information for prognosis that can be explored \cite{gillies2016radiomics}. While many studies have experimented with models that integrate clinical variables, such as size and extent of metastatic involvement \cite{tumorburdenscore, cui2021co}, few have directly applied image analysis on all tumors in a metastatic patient cohort. One particular challenge for such studies is the lack of readily available annotated datasets and standardized theoretical frameworks for studying metastatic and multifocal disease \cite{willemink2020preparing}.

Most existing medical imaging outcome prediction studies focus on only the features of the primary gross tumor volume (GTV) \textit{i.e.} the largest primary tumor when a patient has multiple tumors or metastases. This results in a major oversight as smaller lesions in multifocal tumors or as lymph node metastases may have non-negligible prognostic implications for patients \cite{karaman2014mechanisms}. Chen et al. proposed an autoencoder-based multiple instance learning network based on radiomic features from liver MRI  for studying multifocal cancer outcome prediction for 50 patients and showed that taking features of all lesions into account significantly improved outcome prediction for patients with colorectal cancer liver metastases \cite{chen2021aminn}. A larger-scale study is required to validate the benefits of using information from all tumors for multifocal cancer and metastatic cancer outcome prediction (from here on referred to as multifocal cancer outcome prediction for simplicity), and public datasets are needed to faciliate performance comparisons of network structures across different groups. However, it is difficult to acquire a high-quality metastatic cohort imaging dataset due to the variations in patient management and the lack of detailed clinical annotations including survival information and tumor segmentations.

In this work, we aim to build a benchmarking dataset and theoretical framework for studying the effect of adding additional metastatic tumors for medical-imaging-based outcome prediction. We processed two public datasets, namely HN1 and Lung1 from The Cancer Imaging Archive (TCIA), specifically of patients with multiple tumors, to set up an uniformly processed, richly annotated resource. We also address the lack of expressiveness in commonly-used multiple instance learning functions and propose two alternatives that are more expressive and therefore provide better performance in multifocal outcome prediction. Our experimental results validated that taking imaging features of all tumors and metastases into account outperforms clinical cancer staging and largest-lesion-based approaches and improves outcome prediction in these larger cohorts of HN1-MF (n=83) and Lung1-MF (n=258).

\section{Materials and Methods}
\begin{figure}
    \centering
    \includegraphics[width=\textwidth]{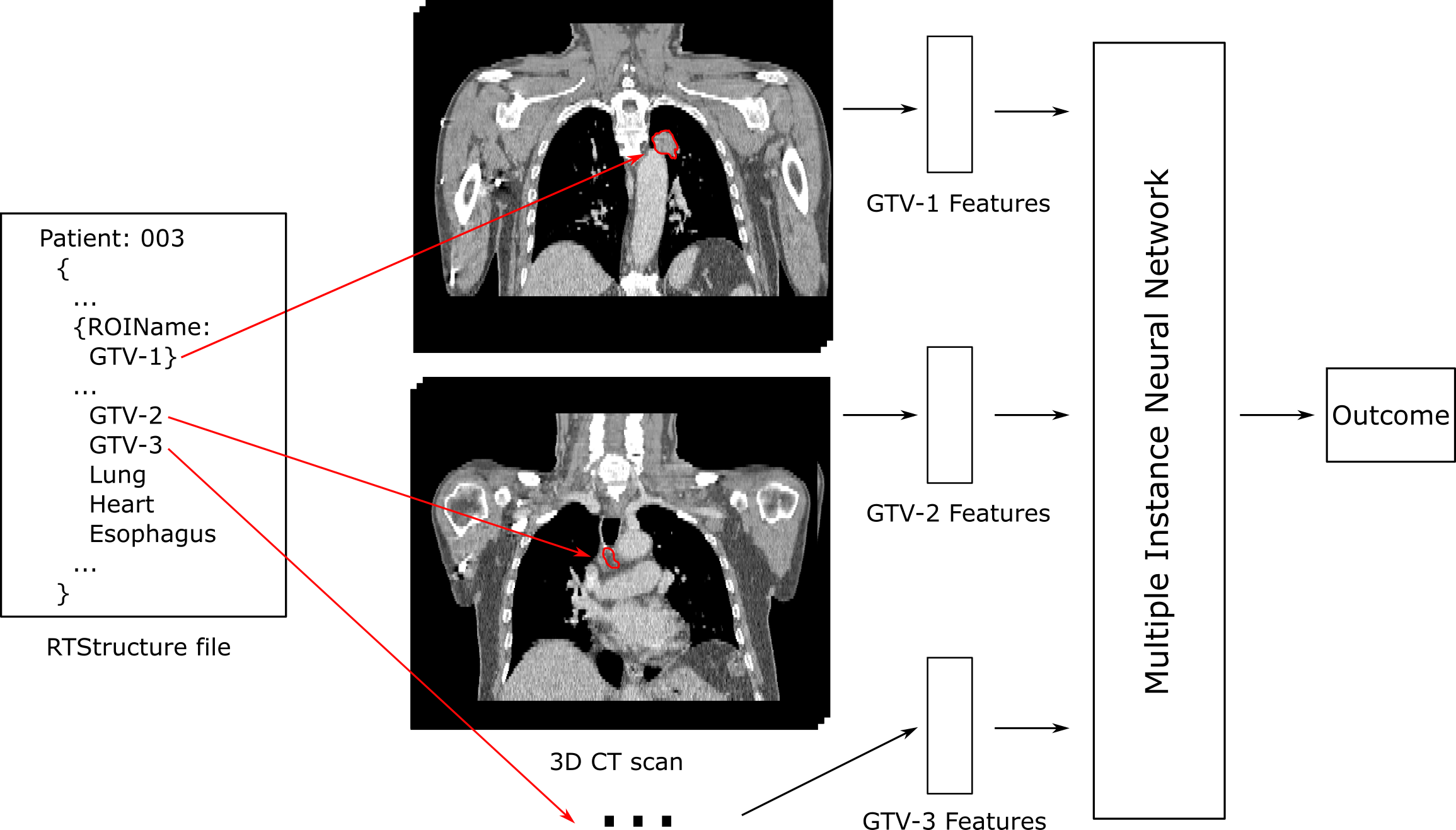}
    \caption{An overview of our study design. GTV annotations are extracted from radiotherapy structure set files to generate masks for all tumors and lymph node metastases in the CT scans. A multiple instance neural network with custom pooling functions predicts patient outcome based on features of all lesions.}
    \label{fig:aminn}
\end{figure}  

The study design is summarized in \textbf{Fig 1}: based on the segmentations of multifocal tumors mined from two public datasets, radiomic features are extracted to represent the characteristics of individual tumors; multiple instance neural network with injective pooling functions are trained to predict patient survival using radiomic features as input. 

\subsection{Multiple Instance Learning}
In contrast to classical supervised learning where a model is trained to predict a target variable $y$ based on a single instance $\textbf{x} \in \mathbb{R}^D$, multiple instance learning (MIL) refers to an architecture where the model predicts the target variable $y$ based on a varying number of $N$ instances from a bag $X=\mathrm{\{\textbf{x}_1, ... \textbf{x}_N\}}$ \cite{carbonneau2018multiple}. MIL models can be trained using only the bag-level label, hence is popular in semi-supervised and self-supervised problems where the instance-level labels are not necessarily available \cite{li2021dual, chen2021aminn, qaiser2021multiple}. In this subsection we discuss the foundations of multifocal cancer outcome prediction as a MIL problem and we propose using \textit{injective} MIL functions, i.e. functions that establish a one-to-one mapping between bag representation and bag contents.

\subsubsection{Problem formulation.}
Multifocal cancer outcome prediction can be posed as a MIL problem as follows: Given a patient with $N$ tumors, we predict the outcome $Y \in \{0,1\}$ of the patient according to the characteristics of the tumors $X=\{\textbf{x}_1, ... \textbf{x}_N\}$. The traditional assumption in MIL is that the bag label $Y$ is negative if and only if all instance labels $y_1, ..., y_N \in \{0,1\}$ are negative \cite{ilse2018attention}:
  \[
    Y=\left\{
                \begin{array}{ll}
                  0, \;\;\;\text{iff} \; \sum_ny_n=0\text{,}\\
                  1, \;\;\;\text{elsewise.}\tag{1}
                \end{array}
              \right.
  \]
For example, a whole-slide image is classified as cancerous if one of its patches appears cancerous, and benign if none of the patches appear so. However, in multifocal cancer outcome prediction, the correlation between instance labels and bag labels is less clear cut. For example, the aggressiveness of a tumor $y_n$ is no longer a binary variable, and it's imprudent to assume that a patient's outcome is determined by a single positive instance. Therefore, we propose the following assumption for multifocal cancer outcome prediction as a MIL problem:
\begin{equation}
    Y= \sum_n y_n \tag{2}
\end{equation}
where the overall risk of a patient $Y \in [0,\infty)$ is the cumulative risk of individual tumors $y_n \in [0,1]$. In the special case of binary survival prediction, as we adopted in this paper, we have:
  \[
    Y_{binary}=\left\{
                \begin{array}{ll}
                  0, \;\;\;\text{iff} \; \sum_ny_n<T\text{,}\\
                  1, \;\;\;\text{elsewise.}\tag{3}
                \end{array}
              \right.
  \]
where $Y_{binary} \in \{0,1\}$ is the binary survival status of a patient, and $T$ is the threshold for a lethal cumulative risk.

\subsubsection{Expressiveness of MIL functions.}
Various MIL pooling methods have been proposed to aggregate instance representations to bag representations \cite{ilse2018attention, wang2018revisiting}. Among them the most popular ones are max, mean and attention-based pooling:
  \[
\mathcal{M} = \left\{
    \begin{array}{ll}
        \text{max}: \textbf{z} = \underset{n}{\mathrm{max}} {\{\textbf{h}_{n}}\} \\
        \text{mean}: \textbf{z} = \frac{1}{N}\sum_{n=1}^{N}\textbf{h}_{n} \\
        \text{attention}: \textbf{z} = \sum_{n=1}^{N}a_{n}\textbf{h}_{n}
    \end{array}
\right. \tag{4}
  \]
where $\textbf{z}$ is the bag representation calculated by aggregating instance representations $\textbf{h}_n$, note that we switched to the embedding-level approach in implementing our MIL model, as our model does not explicitly calculate instance labels $\textbf{y}_n$ but operates at the embedding level \cite{ilse2018attention}. Similar to the readout functions in graph neural networks, we observe that the above MIL pooling functions lacks expressiveness when the instances' effects on bag label is non-binary, because they are not \textit{injective} \cite{xu2018powerful}. \textbf{Fig. 2} demonstrates pairs of bags that confuse max, mean and attention-based pooling. 

\begin{figure}
    \centering
    \includegraphics[width=\textwidth]{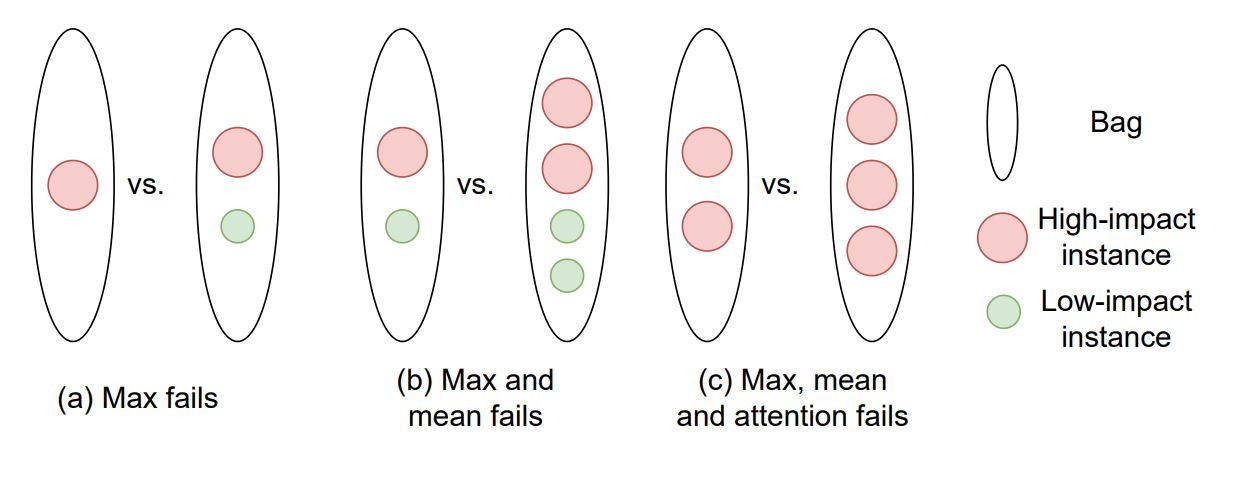}
    \caption{Examples of bags that max, mean and attention-based multiple instance pooling functions fail to distinguish. With our injective functions, the bags in each pair will provide different bag representations.}
    \label{fig:examples}
\end{figure}  

Max pooling only considers the instance with the largest impact to the bag and ignores all other instances (\textbf{Fig. 2a}), which works well with the traditional MIL assumption but may lead to over simplification in a multifocal cancer outcome prediction setting. Let $\textbf{h}_{red}$ and $\textbf{h}_{green}$ denote representations of a high-impact instance and a low-impact instance, respectively. Mean pooling treats the two bags in \textbf{Fig. 2b} as the same since $\frac{1}{2}(\textbf{h}_{red} + \textbf{h}_{green}) = \frac{1}{4}(2 \textbf{h}_{red} + 2  \textbf{h}_{green})$. Attention-based pooling learns instance weight with a multilayer perceptron and softmax activation:
\begin{equation}
    a_n = \frac{\text{exp}(\textbf{w}^\top \text{tanh}(\textbf{Vh}_n^\top)}{\sum_{i=1}^{N}\text{exp}(\textbf{w}^\top \text{tanh}(\textbf{Vh}_i^\top)} \tag{5} \label{eq:attention}
\end{equation}
where \textbf{w} and \textbf{V} are parameters of the network. The learned attentions are normalized by the softmax function based on the instances in each bag. Considering one bag with two red instances and another with three red instances, all traditional MIL pooling generate the same bag representation: $\textbf{h}_{red} = \frac{1}{2} (\textbf{h}_{red} + \textbf{h}_{red}) = ({\frac{1}{2}\textbf{h}_{red} + \frac{1}{2}\textbf{h}_{red}}) = ({\frac{1}{3}\textbf{h}_{red} + \frac{1}{3}\textbf{h}_{red} + \frac{1}{3}\textbf{h}_{red}})$ (\textbf{Fig. 2c}). The attention weights learned for the same type of instances are different in bags with different number of instances, which adds to the difficulty in interpreting the weights. 

\subsubsection{Injective MIL functions.}
In this work, we propose using \textbf{injective} pooling functions to address the differences in MIL assumptions for multifocal cancer outcome prediction. Specifically, we propose two simple MIL functions: sum pooling and unnormalized attention-based pooling:
\begin{equation}
    \mathcal{M}_{\text{sum}}:  \textbf{z} = \sum_{n=1}^{N}\textbf{h}_{n} \tag{6}
\end{equation}
\begin{equation}
    \mathcal{M}_{\text{uatt}}: \textbf{z} = \sum_{n=1}^{N}a_{n}\textbf{h}_{n} \;\;\;\;\;\text{ with } \;\;\;\;\; a_n =\frac{1}{1 + \text{exp}(-\textbf{w}^\top \text{tanh}(\textbf{Vh}_i^\top)} \tag{7}
\end{equation} Sum aggregators can represent universal functions over bags \cite{xu2018powerful}. By simply replacing the softmax function with sigmoid, the attention weights of instances are no longer normalized locally. The injective functions ensure that unique bag representations are learned for each bag, therefore increase the expressive power or multiple instance learning networks. Non-normalized instance representations also paves the way for learning instance risks $y_n$ as continuous variables. Both functions can serve as MIL functions as they are permutation-invariant.

\subsection{Data}
In this retrospective study, we processed two datasets from The Cancer Imaging Archive (TCIA)\footnote{https://www.cancerimagingarchive.net/} and generated new versions of these datasets specifically for multifocal cancer outcome prediction \cite{clark2013cancer}.

The first dataset we used is Head-Neck-Radiomics-HN1 (HN1, version 3)\footnote{https://wiki.cancerimagingarchive.net/display/Public/Head-Neck-PET-CT}, a Computed Tomography (CT) imaging cohort of 137 head and neck squamous cell carcinoma (HNSCC) patients treated by radiotherapy \cite{vallieres2017radiomics}. The second dataset is NSCLC-Radiomics (Lung1, version 4)\footnote{https://wiki.cancerimagingarchive.net/display/Public/NSCLC-Radiomics}, a collection of pretreatment CT scans of 422 non-small cell lung cancer (NSCLC) patients. In both datasets, clinical variables including survival information were collected and 3D gross tumor volumes (GTV) were manually delineated by a radiation oncologist \cite{Aerts2014}.

In each dataset, we investigated the Radiation Therapy Structures (RTStructure) files and selected a subset of patients that had more than one tumors. Patient demographics are presented in \textbf{Table S1}. Academic use of the datasets were granted under the \textit{TCIA No Commercial Limited Access License}.

\subsubsection{Retrieving multifocal segmentations.}
Since the release of HN1 and Lung1, most studies working with these datasets have investigated the correlation between radiomic features of the primary GTV and patient outcome\cite{Aerts2014}. Our analysis differs from previous work by retrieving segmentations of all tumor volumes to predict patient outcome. Specifically, we first read all ROIs in the RTStructure files to extract all the delineated structures. ROIs that consist of tumors of the primary site, nodal metastases and few distant metastases that were selected. In the original RTStructure files, tumors/metastases in the same sites were assigned to the same GTV ROI. In such cases we generated individual tumor masks by detecting disconnected regions and storing them separately. 

\subsection{Feature Extraction}
Radiomic features of each tumor are extracted from CT scans and the corresponding segmentations to serve as inputs for a multiple instance learning neural network \cite{chen2021aminn}. Image preprocessing and radiomic feature extraction were performed following the guidelines of the Imaging Biomarker Standardize Initiative (IBSI)\cite{zwanenburg2020image}. 103 features were extracted. The exact parameters and software we used for feature extraction and a list of the extracted features are reported in \textbf{Table S2}. 

\section{Experiments}
We aim to establish baselines for multifocal cancer outcome prediction for the multifocal patients in Lung1 and HN1. We want to answer two questions: (1) whether injective multiple instance learning (MIL) functions are better than commonly used MIL functions for multifocal cancer outcome prediction, (2) whether we can observe the improvements in prediction performance by taking all tumors into account, in pretreatment Lung CT and Head\&Neck CT. 

In order to have a fair comparison with previous literature, we adopted the same task and evaluation strategy as in \cite{chen2021aminn}. Specifically, we built a multiple instance neural network for predicting right-censored binary survival status of patients in our datasets. Experiments were performed in 10 repeated runs of 10-fold cross validation. We compared our approach with the most popular radiomics model: LASSO logistic regression (LASSO LR) with features of the primary GTV, and AMINN, a state-of-the-art framework for multifocal cancer outcome prediction. We compare the performance of non-injective and injective MIL pooling functions, referred to as `max', `mean', `att', `sum', `uatt' for max, mean, attention-based, sum and unnormalized attention-based pooling, respectively. We also compared the MIL approaches with AJCC clinical cancer staging system, also known as TNM staging\cite{amin2017eighth}.

\subsubsection{Implementation details.}
For our network we follow the architecture and hyperparameters in AMINN, expect that we replace the autoencoder with a 4-layer fully-connected neural network, resulting in a vanilla multiple instance neural network, which is referred to here as `MINN' \cite{wang2018revisiting, chen2021aminn}. Z-score transformation was performed on the radiomic features. The main evaluation metric is area under the receiver operating characteristic (AUC), with 95\% confidence intervals (CI) estimated using the DeLong test based on the average AUCs and covariances in 10 runs \cite{delong}. The baseline logistic regression model is implemented with scikit-learn 0.24.2 and AMINN is implemented using the code released by the authors. Further details can be found in \textbf{Table S3} and our repository that will be released after peer-review.

\section{Results}
\begin{table}[]
    \centering
        \caption{Classification results on HN1-MF in 10 repeated runs of 10-fold cross validation. AUC with 95\% confidence interval is reported. \% of alive patients was calculated at 1 year and 5 years for HN1-MF and Lung1-MF, respectively. *denotes concordance index estimated with cox proportional hazard regression model. }
    \begin{tabular*}{\textwidth}{p{0.22\textwidth}@{\extracolsep{\fill}}lll}
    \hline
        \parbox[t]{2mm}{\multirow{5}*{data}} & Dataset &HN1-MF & Lung1-MF \\
        &\# patients &  83& 258 \\
        &\# tumors & 245&  1039\\
        & Avg \# tumors &3.0& 4.0\\
        & \% survival & 56.6 (5-year)& 63.2 (1-year)\\ \hline
        clinical baseline &AJCC staging & 0.567* (0.494, 0.640) & 0.515* (0.478, 0.552)\\ \hline    
        unifocal baseline &LASSO LR & 0.633 (0.507, 0.759) & 0.559 (0.487, 0.631)\\ \hline
        multifocal baseline&
        AMINN + mean & 0.706 (0.596, 0.816) & 0.583 (0.502, 0.644)\\ \hline
        \parbox[t]{2mm}{\multirow{2}*{non-injective}}&MINN + max  & 0.651 (0.533, 0.769) & 0.569 (0.497, 0.641)\\ 
        &MINN + mean  & 0.748 (0.640, 0.856) & 0.575 (0.502, 0.648)\\ 
        &MINN + att  & 0.723 (0.611, 0.835) & 0.572 (0.499, 0.645) \\ \hline
        \parbox[t]{2mm}{\multirow{2}*{injective}}&MINN + sum  & \textbf{0.773 (0.669, 0.877)} & \textbf{0.596} \textbf{(0.523, 0.669)}\\ 
        &MINN + uatt & \textbf{0.763 (0.657, 0.869)} & $\textbf{0.593} \textbf{(0.519, 0.667)} $\\ \hline
        %\textbf{MI} & \textbf{Attention-based MINN} & \textbf{Volume+Cl} & \bm{$0.672 \pm 0.010$}\\ \hline
    \end{tabular*}
    %Accuracy is presented in the format of ``mean $\pm$ standard deviation''
    \label{tab:results}
\end{table}

245 tumors from 83 patients were extracted from the HN1 dataset to generate the multifocal HN1 dataset (HN1-MF). The median survival time is 2097 days and the models were trained to predict 1-year overall survival (OS) of patients in HN1-MF. 

1039 tumors from 258 patients were extracted from the Lung1 dataset to generate the multifocal Lung1 dataset (Lung1-MF). The median survival time is 523 days and the models are trained to predict 5-year OS of patients in Lung1-MF. Most patients in these two cohorts have advanced cancer stage (\textbf{Table S1}).

We validate our hypothesis by comparing the testing AUC of each method on HN1-MF and Lung1-MF. First, we find that taking more tumors into account with the multiple instance learning setup significantly and consistently improves outcome prediction performance in both datasets, regardless of the choice of model and MIL pooling function. Next, MINN with sum pooling and uatt pooling constantly outperforms commonly used non-injective pooling functions, while the pooling with the least representational power, i.e. max pooling,  gives the worst performance among MIL approaches. In our datasets, AMINN with mean pooling has relatively unstable performance and has worse performance than MINN with mean pooling in the HN1-MF dataset. We observed in our experiments that the simultaneous training of the autoencoder branch and the MIL branch of AMINN is unstable and requires heavy hyperparameter tuning. Surprisingly, AJCC staging, compared to our approach, is not predictive of patient outcome, probably because all the patients in our datasets have more than one tumor and most of them had lymph node involvement (\textbf{Fig. S1}).

\section{Conclusion}
In this paper, we investigated existing public datasets to build two large benchmarking cohorts for multifocal/metastatic cancer outcome prediction. We discussed the difference between the problem formulation of traditional multiple instance learning problems compared to MIL for prognosis. We identified the lack of expressiveness in commonly used MIL pooling functions and showed that the injective MIL functions we proposed are better suited for outcome prediction in multifocal cancer datasets. We also confirmed the hypothesis that taking all primary, nodule and metastatic tumor into account, as opposed to only the largest primary, improves outcome prediction at least in pretreatment non-small-cell lung cancer CT and head and neck cancer CT.

The primary goal of this paper is to establish a reproducible and transparent baseline for developing and benchmarking multifocal/metastatic cancer outcome prediction algorithms. Therefore, we conducted our experiments with binary classification and radiomic features. There are many challenging and interesting questions to pursue. For example, treating survival prediction as a regression problem instead of binary classification \cite{katzman2018deepsurv}, replacing radiomic features with deep learning features \cite{afshar2019handcrafted}, incorporating clinical variables into outcome predictions \cite{wang2021dae}. We believe the resources we have developed and outlined in this study will aid research in these directions and have implications for patient management.

\newpage
\bibliographystyle{splncs04}
\bibliography{IMIL.bib}

\begin{thebibliography}{10}
\providecommand{\url}[1]{\texttt{#1}}
\providecommand{\urlprefix}{URL }
\providecommand{\doi}[1]{https://doi.org/#1}

\bibitem{Aerts2014}
Aerts, H.J., Velazquez, E.R., Leijenaar, R.T., Parmar, C., Grossmann, P.,
  Cavalho, S., Bussink, J., Monshouwer, R., Haibe-Kains, B., Rietveld, D.,
  Hoebers, F., Rietbergen, M.M., Leemans, C.R., Dekker, A., Quackenbush, J.,
  Gillies, R.J., Lambin, P.: {Decoding tumour phenotype by noninvasive imaging
  using a quantitative radiomics approach}. Nature Communications  \textbf{5}
  (2014). \doi{10.1038/ncomms5006}

\bibitem{afshar2019handcrafted}
Afshar, P., Mohammadi, A., Plataniotis, K.N., Oikonomou, A., Benali, H.: From
  handcrafted to deep-learning-based cancer radiomics: challenges and
  opportunities. IEEE Signal Processing Magazine  \textbf{36}(4),  132--160
  (2019)

\bibitem{amin2017eighth}
Amin, M.B., Greene, F.L., Edge, S.B., Compton, C.C., Gershenwald, J.E.,
  Brookland, R.K., Meyer, L., Gress, D.M., Byrd, D.R., Winchester, D.P.: The
  eighth edition ajcc cancer staging manual: continuing to build a bridge from
  a population-based to a more “personalized” approach to cancer staging.
  CA: a cancer journal for clinicians  \textbf{67}(2),  93--99 (2017)

\bibitem{carbonneau2018multiple}
Carbonneau, M.A., Cheplygina, V., Granger, E., Gagnon, G.: Multiple instance
  learning: A survey of problem characteristics and applications. Pattern
  Recognition  \textbf{77},  329--353 (2018)

\bibitem{chen2021aminn}
Chen, J., Cheung, H., Milot, L., Martel, A.L.: Aminn: Autoencoder-based
  multiple instance neural network improves outcome prediction in multifocal
  liver metastases. In: International Conference on Medical Image Computing and
  Computer-Assisted Intervention. pp. 752--761. Springer (2021)

\bibitem{clark2013cancer}
Clark, K., Vendt, B., Smith, K., Freymann, J., Kirby, J., Koppel, P., Moore,
  S., Phillips, S., Maffitt, D., Pringle, M., et~al.: The cancer imaging
  archive (tcia): maintaining and operating a public information repository.
  Journal of digital imaging  \textbf{26}(6),  1045--1057 (2013)

\bibitem{cui2021co}
Cui, H., Xuan, P., Jin, Q., Ding, M., Li, B., Zou, B., Xu, Y., Fan, B., Li, W.,
  Yu, J., et~al.: Co-graph attention reasoning based imaging and clinical
  features integration for lymph node metastasis prediction. In: International
  Conference on Medical Image Computing and Computer-Assisted Intervention. pp.
  657--666. Springer (2021)

\bibitem{delong}
DeLong, E.R., DeLong, D.M., Clarke-Pearson, D.L.: Comparing the areas under two
  or more correlated receiver operating characteristic curves: A nonparametric
  approach. Biometrics  \textbf{44}(3),  837--845 (1988)

\bibitem{dillekaas201990}
Dillek{\aa}s, H., Rogers, M.S., Straume, O.: Are 90\% of deaths from cancer
  caused by metastases? Cancer medicine  \textbf{8}(12),  5574--5576 (2019)

\bibitem{gillies2016radiomics}
Gillies, R.J., Kinahan, P.E., Hricak, H.: Radiomics: images are more than
  pictures, they are data. Radiology  \textbf{278}(2),  563--577 (2016)

\bibitem{ilse2018attention}
Ilse, M., Tomczak, J.M., Welling, M.: Attention-based deep multiple instance
  learning. In: International conference in machine learning (2018)

\bibitem{karaman2014mechanisms}
Karaman, S., Detmar, M., et~al.: Mechanisms of lymphatic metastasis. The
  Journal of clinical investigation  \textbf{124}(3),  922--928 (2014)

\bibitem{katzman2018deepsurv}
Katzman, J.L., Shaham, U., Cloninger, A., Bates, J., Jiang, T., Kluger, Y.:
  Deepsurv: personalized treatment recommender system using a cox proportional
  hazards deep neural network. BMC medical research methodology
  \textbf{18}(1),  1--12 (2018)

\bibitem{li2021dual}
Li, B., Li, Y., Eliceiri, K.W.: Dual-stream multiple instance learning network
  for whole slide image classification with self-supervised contrastive
  learning. In: Proceedings of the IEEE/CVF Conference on Computer Vision and
  Pattern Recognition. pp. 14318--14328 (2021)

\bibitem{qaiser2021multiple}
Qaiser, T., Winzeck, S., Barfoot, T., Barwick, T., Doran, S.J., Kaiser, M.F.,
  Wedlake, L., Tunariu, N., Koh, D.M., Messiou, C., et~al.: Multiple instance
  learning with auxiliary task weighting for multiple myeloma classification.
  In: International Conference on Medical Image Computing and Computer-Assisted
  Intervention. pp. 786--796. Springer (2021)

\bibitem{tumorburdenscore}
Sasaki, K., Morioka, D., Conci, S., Margonis, G.A., Sawada, Y., Ruzzenente, A.,
  Kumamoto, T., Iacono, C., Andreatos, N., Guglielmi, A., et~al.: The tumor
  burden score: a new “metro-ticket” prognostic tool for colorectal liver
  metastases based on tumor size and number of tumors. Annals of surgery
  \textbf{267}(1),  132--141 (2018)

\bibitem{vallieres2017radiomics}
Vallieres, M., Kay-Rivest, E., Perrin, L.J., Liem, X., Furstoss, C., Aerts,
  H.J., Khaouam, N., Nguyen-Tan, P.F., Wang, C.S., Sultanem, K., et~al.:
  Radiomics strategies for risk assessment of tumour failure in head-and-neck
  cancer. Scientific reports  \textbf{7}(1),  1--14 (2017)

\bibitem{wang2021dae}
Wang, C., Sun, X., Zhang, F., Yu, Y., Wang, Y.: Dae-gcn: Identifying
  disease-related features for disease prediction. In: International Conference
  on Medical Image Computing and Computer-Assisted Intervention. pp. 43--52.
  Springer (2021)

\bibitem{wang2018revisiting}
Wang, X., Yan, Y., Tang, P., Bai, X., Liu, W.: Revisiting multiple instance
  neural networks. Pattern Recognition  \textbf{74},  15--24 (2018)

\bibitem{willemink2020preparing}
Willemink, M.J., Koszek, W.A., Hardell, C., Wu, J., Fleischmann, D., Harvey,
  H., Folio, L.R., Summers, R.M., Rubin, D.L., Lungren, M.P.: Preparing medical
  imaging data for machine learning. Radiology  \textbf{295}(1),  4--15 (2020)

\bibitem{xu2018powerful}
Xu, K., Hu, W., Leskovec, J., Jegelka, S.: How powerful are graph neural
  networks? In: International Conference on Learning Representations (ICLR)
  (2019)

\bibitem{zwanenburg2020image}
Zwanenburg, A., Valli{\`e}res, M., Abdalah, M.A., Aerts, H.J., Andrearczyk, V.,
  Apte, A., Ashrafinia, S., Bakas, S., Beukinga, R.J., Boellaard, R., et~al.:
  The image biomarker standardization initiative: standardized quantitative
  radiomics for high-throughput image-based phenotyping. Radiology
  \textbf{295}(2),  328--338 (2020)

\end{thebibliography}
\end{document}